\documentclass[aps,showpacs,floatfix,twocolumn]{revtex4} 
\usepackage{graphicx}
\usepackage{amsmath}
\begin{document}
\title{Low-energy proton capture reactions in the mass region 55-60}
\author{Saumi Dutta}
\email{saumidutta89@gmail.com}
\author{Dipti Chakraborty}
\email{diptichakraborty2011@gmail.com}
\author {G. Gangopadhyay}
\email{ggphy@caluniv.ac.in}
\author{Abhijit Bhattacharyya}
\email{abphy@caluniv.ac.in}
\affiliation{Department of Physics, University of Calcutta\\92, Acharya Prafulla Chandra Road, Kolkata-700 009, India}

\textwidth 7in
\textheight 8in

\begin{abstract}
Low energy proton capture reactions in the mass 55-60 region
are studied in a microscopic optical model. Nuclear density profile is 
calculated using the relativistic mean field theory. The DDM3Y interaction is folded
with the theoretical density to obtain the proton-nucleus optical potential. 
A definite set of normalization parameters has been obtained for the concerned mass region
by comparing with all available experimental data in this mass region.
These parameters have been used to obtain  proton capture rates 
for astrophysically important reactions in this mass region.
\end{abstract}
\pacs{24.10.Ht, 25.40.Cm, 25.40.Lw, 25.40.Kv}
\maketitle

\section{Introduction}

The study of the role of individual nuclear reactions in stellar evolution has 
been an important field 
of research in the last few decades. As a star evolves with time it passes 
through burning in different ranges of nuclear mass. At the same time, 
different nuclear 
processes become important at different time periods of evolution. A 
comprehensive study of these processes sheds light on various astrophysical 
phenomena.  
 
There are certain astrophysical sites which are responsible for the production 
of heavier nuclei beyond iron through the rapid capture of protons on seed
nuclides. In the mass  region of our interest there are certain 
proton rich naturally occurring nuclei, which are not produced by the
$r$-process or the $s$-process. These are called $p$-nuclei. Proton capture reactions in 
certain astrophysical sites can account for the formation of some of these 
proton rich nuclides. For example x-ray bursters with a large proton 
flux in the peak temperature around 1-3 GK are suitable astrophysical sites for 
the production of certain nuclei. To find out the abundance of different 
nuclei  as well as the evolution of the process in these sites a network 
calculation is necessary which involves a large number of reactions. 

It is thus imperative to calculate the rates and/or cross sections of these 
reactions in different mass ranges. Our group has already calculated the 
cross sections and hence the astrophysical S-factors in the mass 
range $A=60-100$~\cite{GG,Lahiri1,Lahiri2,Lahiri3}. Some implications of the new 
rates has also been investigated in the context of rp-process~\cite{Lahiri4,Lahiri5}. 
In the present work, we extend our calculation to the $A=55 - 60$ region. 

The rp-process is sensitive to a  number of reactions in this region.
The most challenging aspect to look at in these scenarios is that most of the 
nuclei involved in those reactions are not produced in the laboratory. 
For example, Parikh {\em et al.}~\cite{parikh} have identified proton capture
reactions on $^{56}$Ni and $^{57,59}$Cu  targets as important in the rp-process
in certain scenarios.
However, experimental rates are not available for these reactions because 
stable targets do not occur in nature. 
Hence, one has to depend on theoretical calculations in this domain. 

In explosive proton rich environments, such as x-ray bursts, proton capture has
to compete with its inverse, {\em i.e.} photo-disintegration. This competition
results in
waiting points and causes delay of further nucleosynthesis. With temperature, 
pressure and proton mass fractions being different at different regions of 
these sites as well as being time-varying quantities, incorporation of all 
these physical conditions in the nuclear network is a big challenge. 
Rauscher {\em et al.}~\cite{Raus1,Raus2} have calculated the 
 rates for various proton, neutron and $\alpha$-particle induced reactions and their reverse reactions 
in Hauser-Feshbach formalism for targets with wide range of atomic numbers and masses and for 
a wide range of temperature.
Theoretical
calculations in this mass region  essentially utilize the Hauser-Feshbach 
formalism where, the optical model potential, a key ingredient, is often 
taken in a local or a global form. However, a more microscopic approach 
is also possible using
an optical potential constructed utilizing nuclear densities. 
If the target is stable, nuclear density is available through electron 
scattering. However, in the absence of a stable target, theory remains our sole 
guide to describing the density. 
It is imperative to test the theoretical 
calculations, where experimental data are available, to verify 
its applicability. We aim to check the success of microscopic optical 
potentials based on mean-field densities in explaining 
the available reaction cross sections in this mass region. A good description 
depending essentially on theory will allow one to extend 
the present method to the critical reactions, which are beyond present day 
laboratory capabilities.

A well 
defined nucleon-nucleon ($NN$) interaction is of major importance for microscopic 
calculation of nucleon-nucleus and nucleus-nucleus potentials used in the 
theoretical analysis of different reactions as well as scattering. The optical 
model potential is highly successful for explanation of different 
branches of nuclear reaction. It can reliably 
predict the basic observables such as total and partial cross sections, elastic 
scattering angular distributions, etc, even for those target nuclei and for 
those energy regions for which no experimental data exist. 
We have used the density dependent M3Y interaction by 
folding the potential with target radial matter densities. 
This interaction has been used in many calculations and has given satisfactory results.

The paper is organized as follows. In the next section, we outline our method 
of calculation. 
Essentially we construct an optical model potential through 
folding an $NN$ interaction with the theoretical density 
profile.
For this purpose we 
use the relativistic mean field (RMF) theory to obtain the density profile of the targets. 
In Sec.~\ref{secresults} the results of our work are discussed in detail. 
Finally we summarize our work.

\section{Model calculation}

The RMF approach has proved to be very successful in 
describing various nuclear properties such as binding energy of nuclei in 
ground states as well as excited states, nuclear density profile, rms charge 
radii, deformation, nuclear halo, moment of inertia, etc~\cite{Ring}. It is 
considered to be the relativistic generalization of the non-relativistic models 
such as Gogny force or Skyrme force Hartree-Fock theory using effective mesonic 
degrees of freedom rather than instantaneous forces. The model is basically 
based upon two major approximations namely mean-field approximation and no-sea 
approximation~\cite {Muller}. The starting point of RMF is 
a suitable Lagrangian density that includes the coupling between the nucleon field 
and meson field as well as meson self couplings so that the 
Lagrangian can successfully describe the properties of finite nuclei as well as the 
equation of state (EOS) of nuclear matter. There are 
different variations of Lagrangian density as well as different 
parameterizations. An accurately calibrated relativistic Lagrangian density, 
FSUGold~\cite{Todd}, has been fitted to the charge radii of nuclei. It contains two 
additional parameters, compared to conventional RMF models, describing self coupling of vector-isoscalar meson and 
coupling between the vector-isovector meson and vector-isoscalar meson. These 
two additional parameters significantly affect the softening of the EOS,
the accurate determination of which is needed for the study 
of various nuclear properties such as charge radii, masses, etc.

Thus theoretical density profiles are extracted in the RMF approach considering 
the FSUGold interaction. The charge density is obtained by convoluting the point 
proton density considering the finite size of the nucleus. 
\begin {equation}
\rho_{ch}({\mathbf r})=e \int \rho( {\mathbf r \prime})g({\mathbf r}-{\mathbf r\prime})d{\mathbf r\prime}
\end {equation}
where $g(r)$ is the Gaussian form factor given by,
\begin{equation}
g(r)=(a\sqrt{\pi})^{-3} exp(-r^{2}/a^{2})
\end{equation}
where $a$ is a constant whose value is assigned to 0.8 fm.
Using the nuclear density profile we have numerically obtained rms charge radii.

While calculating the charge density or the radius, no attempt has been made to
take the correction due to center of mass into account. Calculations on 
harmonic oscillator wave functions show that the correction is small for heavier
nuclei. For example, Quentin  has shown\cite{Quentin} that the effect of inclusion of center 
of 
mass correction in the radius is given by $\delta r/r\approx 0.9/A^{4/3}$. 
Hence,  we do not expect the density profile to be affected significantly 
due to this approximation.


The M3Y interaction~\cite{M1,M2} is based on a realistic G-matrix which in turn is 
constructed in a harmonic oscillator representation averaging over a range of energies 
as well as densities. It has no explicit density dependence nor energy 
dependence. Although in most cases these averages do not matter 
producing satisfactory results, in few cases it becomes necessary to 
incorporate explicit density dependence into M3Y interaction and then it is 
named as density dependent M3Y (DDM3Y) effective interaction~\cite{DM}. 
Low energy proton capture reactions are highly sensitive to nuclear radius as 
well as density. In the present work we have used density dependent M3Y 
Reid-Elliot effective nucleon-nucleon interaction  within a folding model
prescription \cite{Lahiri1}. The density dependence is incorporated in the same way as 
suggested in Refs.~\cite{DN1,DN2}.

Further, we have included a spin-orbit term into the potential considering Scheerbaum 
prescription~\cite{Scheer} which has been coupled with the phenomenological complex potential 
depths. These depths are functions of energy which are assigned standard values as in 
Lahiri {\em et al}.~\cite{Lahiri1}. These values are kept unaltered throughout our present work.

We have incorporated the density dependent M3Y interaction within the TALYS1.4 
code~\cite{Talys} and performed a Hauser-Feshbach (HF) calculation. 
We have chosen Goriely's microscopic level densities and 
Hartree Fock Bogolyubov model for $E1$ $\gamma$-ray strength function.
As seen in our previous calculations~\cite{Lahiri1,Lahiri2}, these choices can 
explain the experimental results more accurately. All these options are 
available in the code. We have also included the effect of the width fluctuation
correction which has a significant impact at low incident energies. Up to
 30 discrete levels are included for both target and residual nuclei, which
are considered in Hauser-Feshbach decay and $\gamma$-ray cascade. We also include 
a maximum of 30 discrete levels for the nuclei resulting from binary emission in
Hauser-Feshbach decay and $\gamma$-ray cascade. HF calculations are 
done with full $j,l$ coupling.
We have incorporated the density data obtained from RMF approach to obtain the 
optical model potential.

Because of rapid variation of cross-section with energy in the low energy 
region, it is difficult to compare the theory and experiment. A standard 
alternative way is to compare another important quantity instead of 
cross-section, namely astrophysical S-factor \cite{Rolfs}.

The proton capture reactions in astrophysical sites occur within a narrow energy window~\cite{Rolfs}. This effective energy window approximately of Gaussian shape around a peak (known as Gamow peak) is known as Gamow window. 
The expressions for the Gamow peak and Gamow width in a practical form are given respectively as,
\begin{equation}
E_{0}=0.1220(Z_{t}^{2}Z_{p}^{2}\mu T_{9}^{2})^{1/3} MeV
\end{equation}
\begin{equation}
\Delta = 0.2368 (Z_{p}^{2}Z_{t}^{2}\mu T_{9}^{5})^{1/6}
\end{equation}
wherei $\mu$ is the reduced mass and $T_{9}$ denotes the temperature in GK. 
Thus most of the astrophysically important reactions occur within a narrow energy 
window $E_{0}-{\Delta}/{2}$ to $E_{0}+{\Delta}/{2}$.
We see that for ($p,\gamma$) reactions on stable isotopes in the mass range 
55-60, the Gamow window lies  between 1 MeV to 3 MeV for temperature around 
3 GK. Hence, we have carried out our calculation in this low energy window 
and compared our results with the measured data where available. In calculating the Gamow 
peak, Gamow width and hence Gamow window we have taken the masses from Audi 
{\em et al}.~\cite{Audi}.

\section{Results}
\label{secresults}
\begin{figure*}[htp]
\includegraphics[scale=0.475]{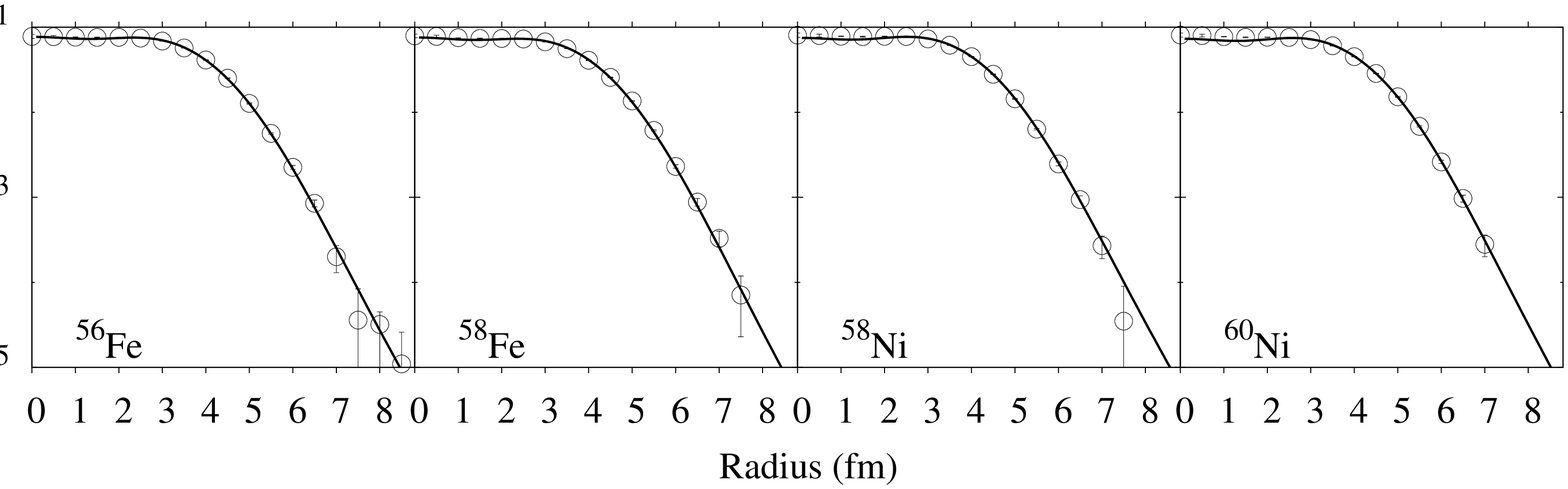}
\caption{\footnotesize Theoretical density profile of various nuclei in mass 
55-60 region from RMF approach compared with experimental data obtained by 
elastic electron scattering taken from Wohlfahrt et. al.~\cite{Wohl}. The solid 
lines denote the theoretical result and discrete points with error-bars 
represent the experimental data. In most of the cases, especially for lower 
radii values, the errors associated are smaller than the dimension of the empty 
circles.}
\label{fig:den}
\end{figure*}

Because the optical model is dependent on the density profile of the nucleus, we calculate 
the density and the charge radii of nuclei in this mass region using RMF formalism. The 
theoretical  density values are plotted as a function of radius and compared with available 
experimental values in Fig.~\ref{fig:den}. As can be seen the agreement is extremely good. 
The experimental data are taken from Wohlfahrt {\em et al}~\cite{Wohl}. 

We also compare all the available rms charge radii values with theoretical results in 
table~\ref{tab:exp2}. The experimental values are taken from Angeli {\em et al}.~\cite{Ang}. 
It can be seen that the RMF calculation has an excellent predictive power, 
the relative difference between theory and experiment in all cases being less than 0.5\%. 
\begin{table}[htp]
\center
\caption{\footnotesize Charge radii of various nuclei extracted in the RMF approach 
compared with measured values from Angeli {\em et al}.~\cite{Ang}}
\label{tab:exp2}
\begin{tabular}{ c c c c c} 
\hline
Nucleus     &   \multicolumn{2}{c} {Charge radius(fm)} \\
  &                Theory   &      Experiment.   \\
\hline
$^{55}$Mn  &     3.7057 &         3.7057 \\
$^{56}$Mn  &     3.7189 &         3.7146 \\
$^{56}$Fe  &     3.7361 &         3.7377 \\
$^{57}$Fe  &     3.7497 &         3.7532 \\
$^{58}$Fe  &     3.7634 &         3.7745 \\
$^{59}$Co  &     3.7924 &         3.7875 \\
$^{58}$Ni  &     3.7916 &         3.7757 \\
$^{60}$Ni  &     3.8193 &         3.8118 \\
\hline
\end{tabular}
\end{table}\\

\begin {figure*}[htp]
\includegraphics[scale=0.45]{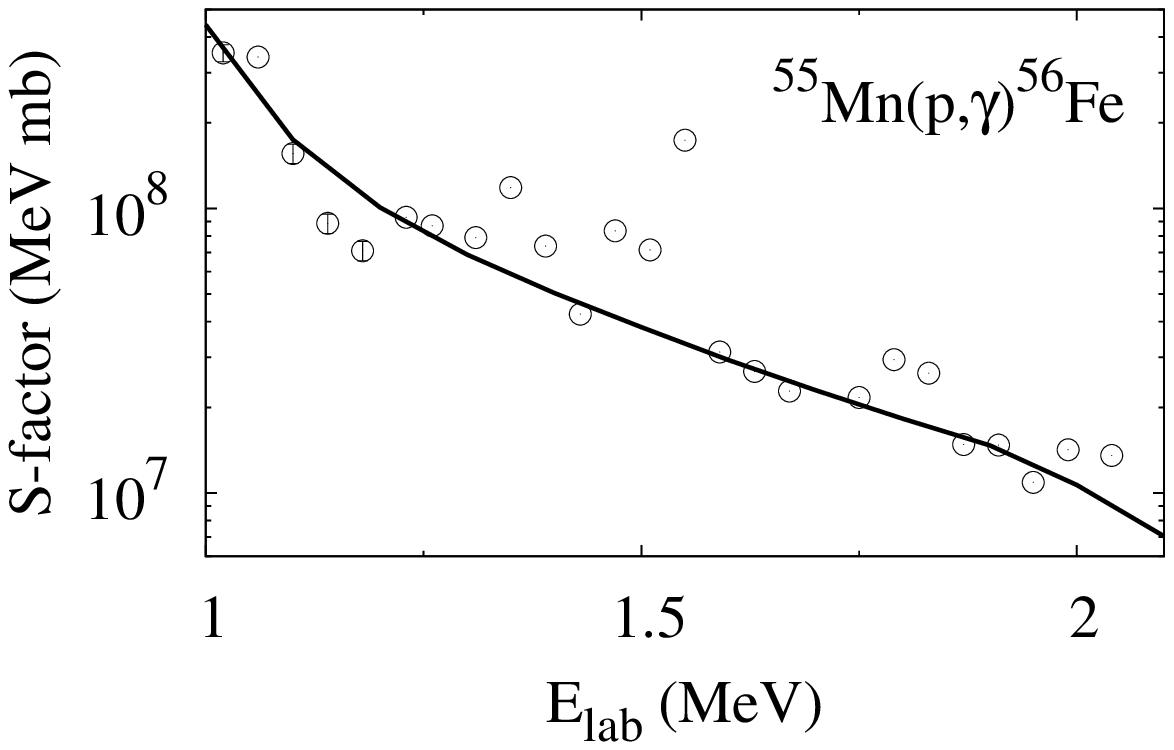}
\includegraphics[scale=0.45]{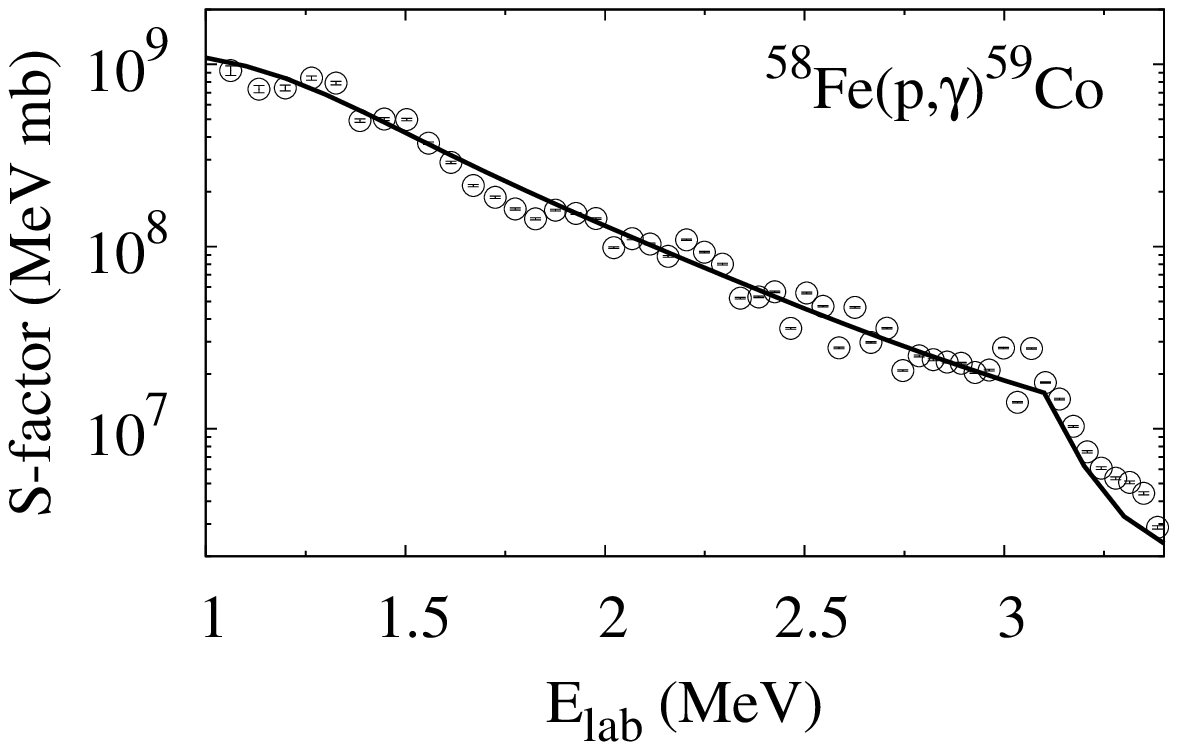}
\includegraphics[scale=0.45]{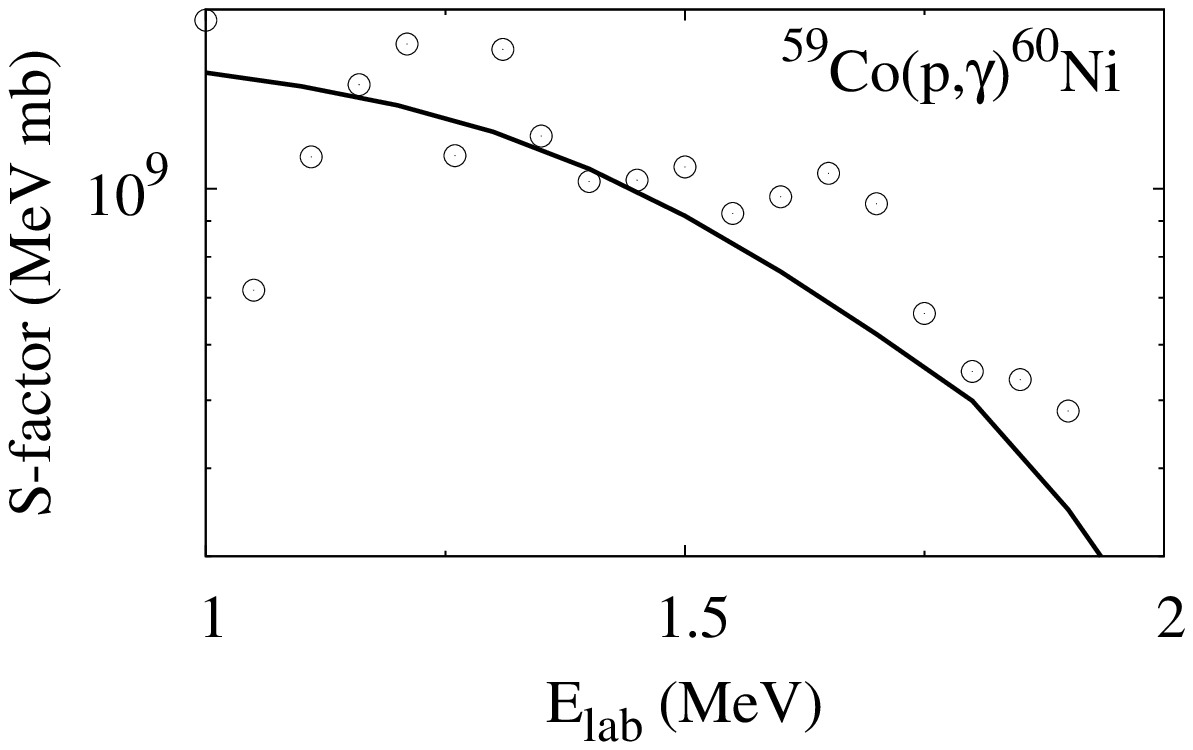}
\includegraphics[scale=0.45]{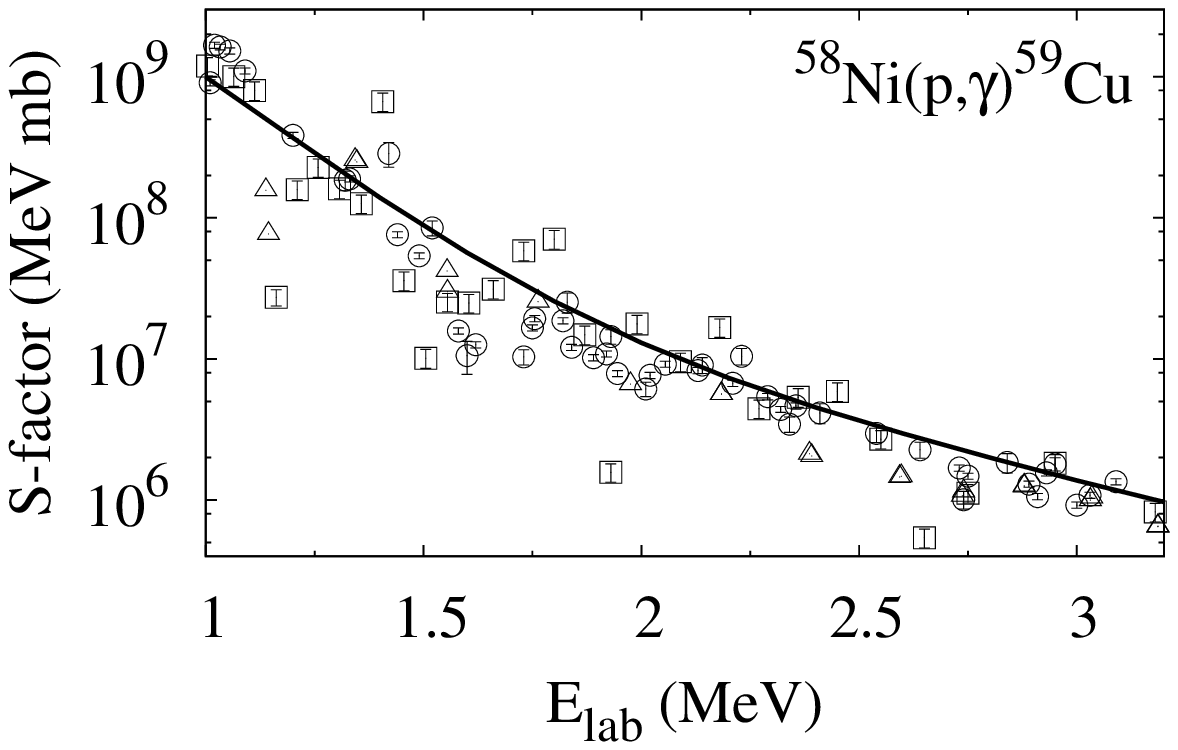}
\includegraphics[scale=0.45]{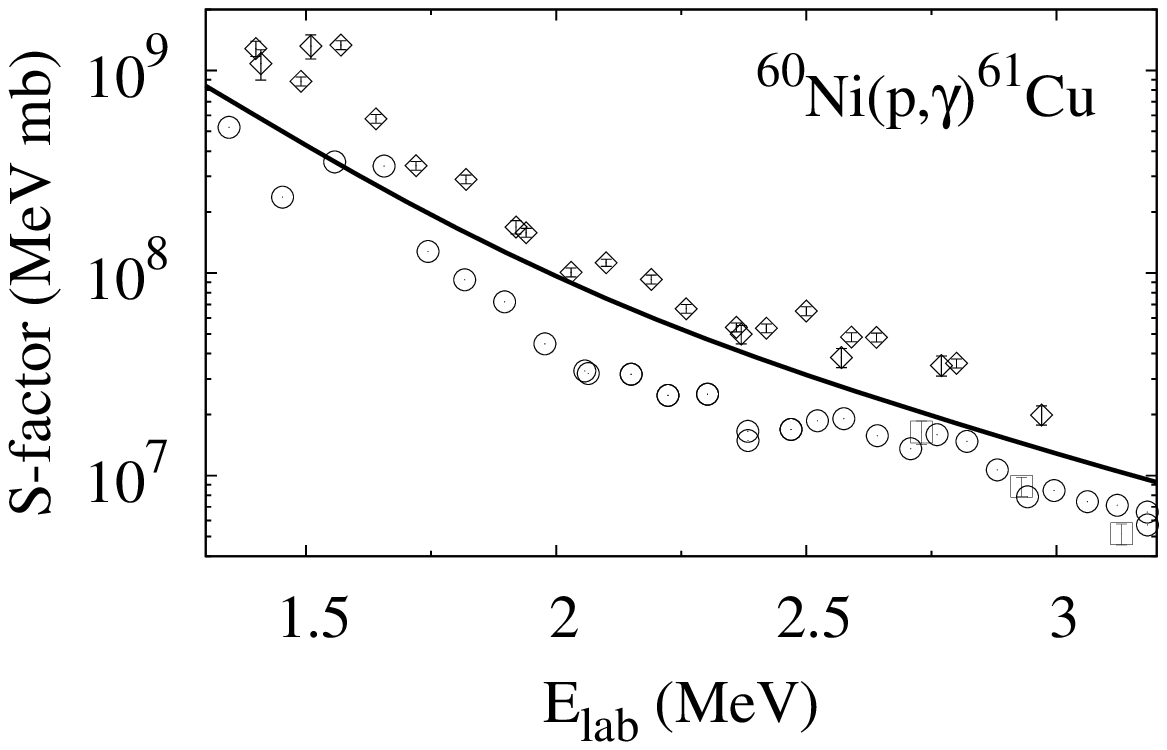}
\caption{\footnotesize Comparison of theoretical astrophysical S-factors
calculations in the present work with experimental data for proton capture 
on (a)$^{55}$Mn, (b)$^{58}$Fe, (c)$^{59}$Co, and (d)$^{58}$Ni, (e)$^{60}$Ni. 
See text for the explanation of different symbols and the Supplemental Material in 
Ref.~\cite{supply} for numerical values of S-factor.}
\label{fig:ddm3y} 
\end{figure*}
We have tried to set a definite normalization for the optical model potential 
that fits all the reaction data in the concerned mass region. The potential 
obtained by folding has been multiplied by the normalization constant 2.0 to 
get the real part of the potential. The DDM3Y interaction does not have any imaginary 
part. We have multiplied the folding potential by the normalization 
constant 1.4 to obtain the imaginary part of the optical potential. 
These final parameters have been obtained after many trials to ensure a 
reasonable agreement with experimental data for all the known low energy proton 
capture reactions in the mass region of our interest. 
Although a single 
normalization can not reproduce the experimental data excellently for all 
reactions in the region, {\em i.e.} each individual reaction may have different 
normalization for best matching with measurement, it is necessary to consider a 
single definite normalization to extend the work to unknown nuclei in the mass 
region for which no experimental data exist. 
We note that the fitted parameters for the present mass range differ
from the neighbouring mass region in our earlier calculation. This is possibly 
due to the fact that the mass selected in the present calculation is lighter
than our previous regions.  
Possibly, the larger depths of the potential is required  to adjust for the low 
mass region.

The comparison of S-factors obtained after incorporating DDM3Y interaction using
the above normalization constants with experimental data are shown in fig. 
\ref{fig:ddm3y}. The numerical values of the S-factors and the reaction rates are 
given in Supplemental Material~\cite{supply}. The experimental data are taken from Ref~\cite{Mn,Fe,Co} for 
$^{55}$Mn, $^{58}$Fe and $^{59}$Co, respectively. For $^{58}$Ni and $^{60}$Ni,
 experimental data are taken from Refs.~\cite{Ni581,Ni582,Ni583} and 
 Refs.~\cite{Ni601,Ni602,Ni582}, respectively. 
In many cases the experimental data are very old.  Errors are also not 
available in some cases. For $^{55}$Mn, $^{58}$Fe and $^{59}$Co,
circles represent the experimental data. For $^{58}$Ni, triangles, 
squares and circles represent the data from Refs.~\cite{Ni583,Ni581,Ni582}, respectively. 
For $^{60}$Ni there are three different sources of data~\cite{Ni601,Ni602,Ni582} which are denoted by 
squares, circles and diamonds, respectively.

\begin{figure*}[hbtp]
\includegraphics[scale=0.458]{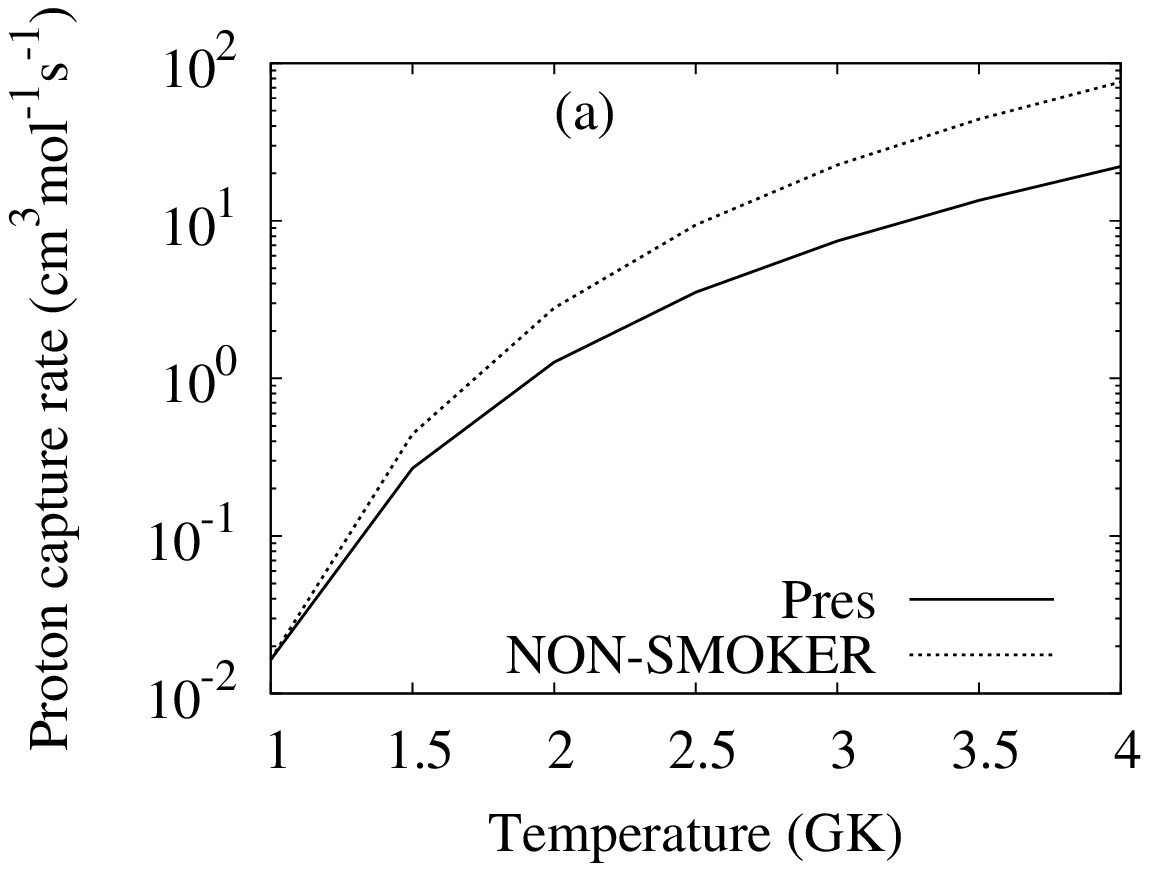}
\includegraphics[scale=0.458]{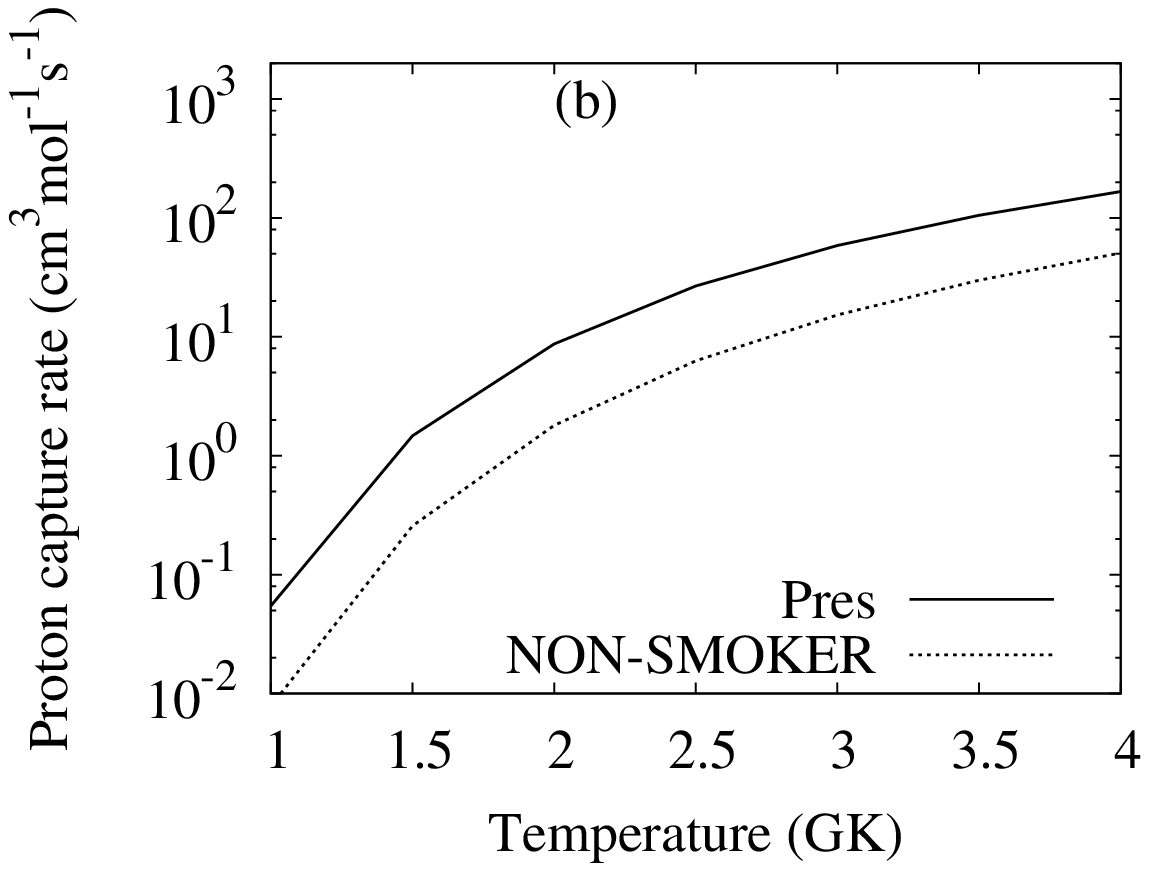}
\includegraphics[scale=0.458]{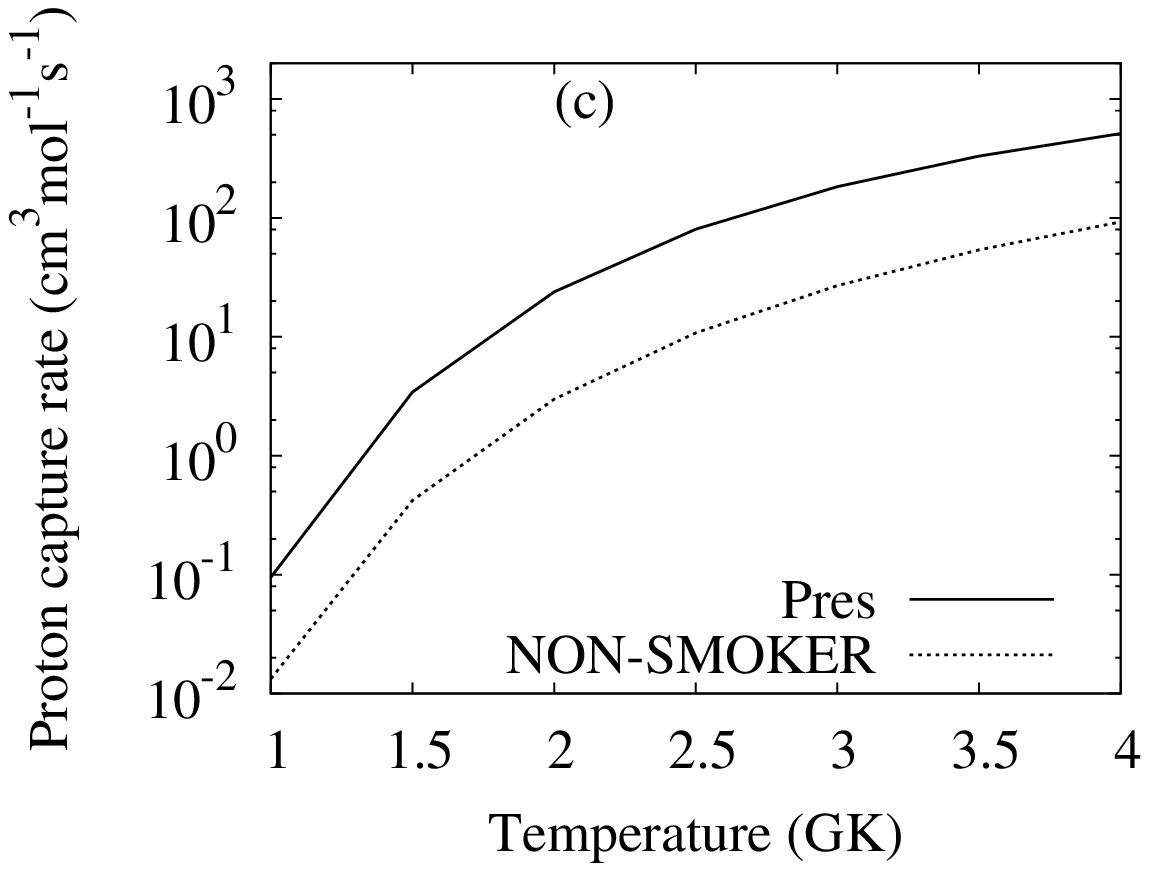}
\caption{\footnotesize Comparison of astrophysical rates from the present calculation 
with NON-SMOKER results from Rauscher {\em at al.}~\cite{Raus1,Raus2} for the 
reactions (a)$^{56}$Ni($p,\gamma)^{57}$Cu, 
(b) $^{57}$Cu($p,\gamma)^{58}$Zn and (c) $^{59}$Cu($p,\gamma)^{60}$Zn.
See text for details \label{fig:rate} and the Supplemental Material in 
Ref.~\cite{supply} for numerical values of rates.}  
\end{figure*}

In all cases the solid line denotes the theoretical DDM3Y result.
In $^{55}$Mn, there are certain ambiguities in 
experimental data, especially in the energy range between 1.3 to 1.6 MeV. The 
experiment was done using Ge(Li) detector by integrated beam current method
more than three decades ago. However, errors are not associated with most of the
data points. Only four data points in the energy range of our interest have 
errors associated with those. 

Our calculations give an excellent description of 
 experimental data for $^{58}$Fe. The experiment for $^{58}$Fe was done using 
Ge(Li) detector and the data was compared with statistical model predictions~\cite{Fe}. 
For $^{59}$Co again there are large fluctuations in experimental 
data. Butler {\em et al.}~\cite{Co} stated that they had observed several resonances 
in the reaction $^{59}$Co$(p,\gamma)^{60}$Ni but the resonances were too close 
to be resolved clearly. 

Our calculation for $^{58}$Ni overpredicts the measurement of Tingwell {\em et al.}~\cite{Ni583} 
by a factor of $\sim$ 2.5, approximately. This experiment was 
carried out by both beam current integrated method and single target irradiation method using Ge(Li) detector. 
Tingwell {\em et al.} also compared their data with 
statistical model calculations. They found that their statistical calculation 
overestimates the measurement by a factor of $\sim$ 2.5 for $^{58}$Ni,  which 
agrees with our results. Cheng {\em et al.}~\cite{Ni581} also measured cross section 
for this reaction using the activation technique. Except in the energy range 
$\sim$ 1.4-1.8 MeV, where the measurement itself has large discrepancies, the 
data agree more or less well with our theoretical calculations. 

For the reaction $^{60}$Ni$(p,\gamma)^{61}$Cu,  Tingwell {\em et al.} themselves
compared the experimental results with the statistical model predictions and 
showed that normalizing the optical model imaginary well depth for Ni isotopes 
by a factor of 1.5 leads to a better agreement between theory and experiment~\cite{Ni602}. 
Our calculation, in the case of $^{60}$Ni, overpredicts the 
experimental  data of Tingwell {\em et al.}~\cite{Ni602} by a factor $\sim$ 1.5,
whereas it underpredicts the  data of Krivonosov {\em et al.}~\cite{Ni582} by a factor $\sim$ 0.35.

With the above normalization, we have calculated the rates for $(p,\gamma)$ 
reactions identified as important by Parikh {\em et al}.~\cite{parikh}. 
The calculated rates are compared with NON-SMOKER~\cite{Raus1,Raus2,Raus3} rates.
The NON-SMOKER results are from a HF calculation based on masses from 
experimental measurements and calculation in the Finite Range 
Droplet Model~\cite{FRDM}. Other details of the calculation can be obtained 
from the references. The results have been plotted in figure 3. 
We see that in the range 1-4 GK, the NON-SMOKER results differ 
from the present calculations significantly.  For the $^{56}$Ni($p,\gamma)$ reaction, although 
the results agree at low temperature, at higher temperature the NON-SMOKER
rates are larger. On the other hand, for the other two reactions, {\em viz.}
$^{57}$Cu($p,\gamma)$ and $^{59}$Cu($p,\gamma)$, our calculation
predicts a significantly larger rate throughout the temperature range.
It will be interesting to see the effects of these results on 
astrophysical scenarios.

\section{Summary}

Low energy $(p,\gamma)$ reactions are studied in a semi-microscopic approach 
in the HF formalism and compared with experiments in the mass region 55-60. 
Radial density profiles are obtained using the RMF approach and are folded
with the DDM3Y NN interaction to obtain semi-microscopic optical potentials. 
Both the real and imaginary depths of the potential are normalized to obtain a 
good agreement between theory and experiment. The S-factors for $(p,\gamma)$
reactions are evaluated in the Gamow window corresponding to 3 GK. We have 
not modified the parameters to fit individual reactions as our aim is to 
construct a framework  for calculation of astrophysical reactions involving 
unstable nuclei. Rates for important astrophysical reactions calculated
in the present approach differ significantly from NON-SMOKER rates. The key 
feature of our work is that we have taken all nuclei 
in the same footing and same methodology has been used for all of them to avoid systematic error. 

\section*{Acknowledgment}
The authors acknowledge the financial support provided by UGC(DRS), DST and the 
University of Calcutta.

\end{document}